# Importance of interactions for the band structure of the topological Dirac semimetal Na$_3$Bi

I. Di Bernardo,[1,2] J. Collins,[1,2] W. Wu,[3] Ju Zhou,[4] Shengyuan A. Yang,[3] Sheng Ju,[4] M. T. Edmonds,[1,2,5*] M. S. Fuhrer[1,2,5*]

[1] Australian Research Council Centre of Excellence in Future Low-Energy Electronics Technologies, Monash University, 3800 Clayton, Victoria, Australia

[2] School of Physics and Astronomy, Monash University, 3800 Clayton, Victoria, Australia

[3] Research Laboratory for Quantum Materials, Singapore University of Technology and Design, Singapore, 487372, Singapore

[4] School of Physical Science and Technology, Soochow University, Suzhou, 215006, China

[5] Monash Centre for Atomically Thin Materials, Monash University, 3800 Clayton, Victoria, Australia

* Corresponding authors: mark.edmonds@monash.edu and michael.fuhrer@monash.edu

**Abstract**

We experimentally measure the band dispersions of topological Dirac semi-metal Na$_3$Bi using Fourier-transform scanning tunnelling spectroscopy to image quasiparticle interference (QPI) on the (001) surface of molecular beam epitaxy-grown Na$_3$Bi thin films. We find that the velocities for the lowest-lying conduction and valence bands are $1.6 x 10^6\ ms^{-1}$ and $4.2 x 10^5\ ms^{-1}$ respectively, significantly higher than previous theoretical predictions. We compare the experimental band dispersions to the theoretical band structures calculated using an increasing hierarchy of approximations of self-energy corrections due to interactions: GGA, meta-GGA, HSE06 and GW methods. We find that density functional theory methods generally underestimate the electron velocities. However, we find significantly improved agreement with an increasingly sophisticated description of the exchange and interaction potential, culminating in reasonable agreement with experiments obtained by GW method. The results indicate that exchange-correlation effects are important in determining the electronic structure of this Na$_3$Bi, and are likely the origin of the high velocity. The electron velocity is consistent with recent experiments on ultrathin Na$_3$Bi and also may explain the ultra-high carrier mobility observed in heavily electron-doped Na$_3$Bi.

**INTRODUCTION**

Topological Dirac semi-metals (TDS) such as Na$_3$Bi and Cd$_3$As$_2$ are a new class of materials featuring two degenerate Weyl points with opposite topological charge occurring at the same momentum protected by time reversal symmetry and rotation and inversion.[1–4] TDS materials have yielded exciting physical properties, such as ultra-low potential fluctuations comparable to the highest quality graphene on h-BN,[5] the chiral anomaly,[6,7] Shubnikov de-Haas oscillations consistent with Weyl orbits,[8] weak anti-localization with phase coherence lengths of up to 1 μm,[9] Dirac gap dependence on the film thickness,[10] and an electric field induced transition from topological insulator to conventional insulator in ultra-thin $Na_3Bi$.[11]

The low-energy bandstructure of $Na_3Bi$ results from the overlap of the conduction band dominated by Na-3$s$ states and the valence band dominated by Bi-6$p$ states. Hybridization gaps most of the overlap region except for the two gapless Dirac points which are protected by $C_3$ rotational symmetry. The Dirac points are thought to be distinct only over a region of a few tens of meV, undergoing a merger (Liftshitz transition) at higher energies.[12] Interestingly, large variations in mobility are observed between lightly and heavily n-type doped $Na_3Bi$. For $Na_3Bi(001)$ bulk crystals[1,7,13–15] and thin-films grown on insulating $Al_2O_3[0001]$,[9,16,17] where the Fermi energy lies in the regime of distinct Dirac points, between 10-40 meV above the Dirac node (i.e. n-type doped) the electron mobility is between 1000-8000 $cm^2V^{-1}s^{-1}$. In this regime the bands have mainly Bi-6$p$ character. In contrast, heavily n-type doped $Na_3Bi$ bulk crystals, where the Fermi energy is hundreds of meV above the Dirac node, i.e. above the Lifshitz transition and well into the conduction band with stronger Na-3$s$ and Bi-$s/p_z$ character, possess electron mobilities that can exceed 100 000 $cm^2V^{-1}s^{-1}$.[18] This suggests a strong asymmetry in the electron (Na-3$s$ and Bi-$s/p_z$) and hole (Bi-6$p$) band velocities. Angle-resolved photoelectron spectroscopy has provided evidence of linearly dispersing bulk and surface bands below or near to the Dirac node on both bulk and thin-film $Na_3Bi(001)$.[1,11,15] Yet, little is known about the band dispersion in the conduction band well above the Dirac node (and above the Liftshitz transition energy).

Fourier-transform scanning tunnelling spectroscopy[19–25] or quasiparticle interference (QPI) offers a powerful means to study the dispersions of both filled and empty bands. Quasiparticles arising from the bulk or the surface of the sample scatter from surface potential barriers caused by lattice imperfections or defects, resulting in a standing wave interference patterns of characteristic wavevector $\mathbf{q} = \mathbf{k_i} - \mathbf{k_f}$, where $\mathbf{k_i}$ and $\mathbf{k_f}$ are the initial and final momenta of the scattered wave. This modulation in the local density of states induced by the quasiparticle interference can then be mapped locally via scanning tunnelling spectroscopy, where the Fourier transform of the differential conductance map (FT-STS) is approximately characterized by wavevectors that connect the high density of states regions in the band structure. Hence, QPI patterns resemble the joint density of states (JDOS) for the surface state electrons, similar to that obtained with ARPES measurements.

Here we use QPI to study large-area, atomically flat $Na_3Bi(001)$ thin films grown on Si(111) via molecular beam epitaxy using low temperature scanning tunnelling microscopy and spectroscopy. The voltage-dependent quasi-particle interference mapping of the surface local density of states is measured at positive and negative bias in order to map the Dirac-like dispersion of the electron and hole bands, and ring-like structures, with wavevector scaling linearly with energy, are observed in the FT-STS, and identified as the expected features at $\mathbf{q} = 2\mathbf{k_f}$ for a circular constant-energy surfaces. We find significant differences between the experimentally determined band dispersion and previous theoretical modelling.[3] In particular, we observe a very high velocity ($1.6 \times 10^6\ ms^{-1}$) for the electron band, which is approximately three times higher than previously anticipated theoretically, but is supported by the observation of a high-velocity conduction band via ARPES in ultrathin Na$_3$Bi.[11]

We compare the experimental band velocities to the theoretical band structure calculated using *ab initio* theory with an increasing hierarchy of approximations for the interaction corrections: DFT, i.e., GGA, SCAN (meta-GGA), HSE06, and GW method. We find that the band structure depends strongly on the level of approximation for interaction corrections. Our GGA calculation agrees with previous results[3] but underestimates the conduction band velocity by 69%, while the GW calculation obtains excellent agreement for the conduction band velocity. The significant increased velocity in HSE06 compared to GGA, and the good agreement obtained between theory and experiment with GW method, suggests that exchange-correlation effects are possibly at the origin of the high velocity.

**Experimental Methods:**

Experiments were performed in a combined low-temperature scanning tunnelling microscope (STM) that operates at 4.7 K in ultrahigh vacuum equipped with an interconnected molecular beam epitaxy chamber (MBE). 20 nm Na$_3$Bi thin film were grown on Si(111) 7x7 in the MBE chamber according to the procedure established in ref. [5, 17]: a Si(111) wafer was flash-annealed to achieve an atomically flat 7×7 surface reconstruction. The substrate temperature was kept at 330°C during growth and Na and Bi were co-deposited to achieve a 20 nm thickness. After growth the sample was left at 330°C for further 10 min in Na overflux to improve crystallinity. Na and Bi rates were calibrated with a quartz crystal monitor. A PtIr STM tip was prepared and calibrated using an Au(111) single crystal and the Shockley surface state before all measurements. STM differential conductance (d$I$/d$V$) was measured using a 10 mV AC excitation voltage (673 Hz) that was added to the tunnelling bias. All STM measurement were collected at 4.7 K. Differential conductance maps and 2D Fourier transform were analysed using WSxM and IGOR Pro softwares.

The first-principles calculations were based on the density functional theory (DFT), as implemented in the Vienna ab initio Simulation Package (VASP).[26,27] The projector augmented wave (PAW) method was used for treating the ionic interaction.[28] The generalized gradient approximation (GGA) with Perdew-Burke-Ernzerhof (PBE) realization were adopted for the exchange-correlation (XC) potential.[29] The plane-wave cutoff energy was set to be 400 eV and the Brillouin zone was sampled with a $10x10x6\ \Gamma$-centered $k$ mesh. The structure was fully optimized with the energy and force convergence criteria $10^{-6}$ eV and $10^{-2}$ eV/Å, respectively. The optimized van der Waals (vdW) correlation functional optB86b-vdW was included.[30] In addition to the GGA method, band structure was also calculated by the meta-GGA SCAN method,[31,32] the more accurate hybrid functional (HSE06) method[33,34] and the GW approach.[35] For the GW calculation, the Brillouin zone was sampled with a

8x8x4 $\Gamma$-centered $k$ mesh. To ensure the sufficient empty bands, the total number of bands is set to be 10 times the number of valence bands.

**RESULTS AND DISCUSSION**

The overall quality of the 20 nm Na$_3$Bi films grown on Si(111) is verified via scanning tunnelling microscopy and spectroscopy data. Figure 1(a) shows a typical large scale (300 nm x 300 nm) topographic STM image of the sample, characterized by relatively flat, smooth terraces as large as 100 nm. In the inset of Fig. 1(a) we report an atomically resolved image of a separate film grown under the same experimental conditions, exhibiting the (1x1) Na-terminated surface with a lattice constant of 5.45Å (marked by white arrows in the inset) in good agreement with previous experimental works and theoretical predictions.[9,10]

Figure 1(b) shows the STS spectrum acquired on a single point on the film in a [-550 mV; +500 mV] range: the local density of states (LDOS) exhibits a clear minimum - identified with the apex of the Dirac cone $E_D$ - about 10 meV above the Fermi level, indicating that the film is slightly $p$-doped, in line with previous reports.[5] About 40 meV below $E_D$ we observe the typical resonant feature [labelled D in the inset of Figure 1(b)] caused by the presence of the Na vacancies on the surface of Na$_3$Bi.[5] The overall lower intensity of the conduction band compared to the valence band implies a lower number of available states for the electrons above the Dirac cone, hence a lower number of available bands in this energy range. The spectrum also exhibits sharp increases of the density of states at about -350 and -480 meV: these onsets are attributed to maxima of valence bands.

In order to accurately probe the electronic band structure of Na$_3$Bi above and below the Dirac node with FT-STS high $q$-space resolution is required, which due to the large lattice constant (0.55 nm) necessitated performing differential conductance maps on topographic regions larger than 55 nm. Figure 2(a) shows an STM topographic image (60 nm x 60 nm) of an atomically flat region of Na$_3$Bi(001) grown on Si(111). A large number of defects corresponding to Na surface vacancies are observed, appearing as darker spots. These vacancies are responsible for the overall p-type doping of the Na$_3$Bi film, and contribute to creation of electron- and hole-like charge puddles.[5] These vacancies also act as a perturbation to the surface potential, consequently acting as scattering centres for the Dirac fermions in Na$_3$Bi, leading to the emergence of quasiparticle interference. Our QPI maps were performed on regions free from any multi-vacancies (i.e. non-singular defect sites) such as Figure 2(a), due to these types of vacancies giving rise to tip-induced charge ionization rings[5] that obscure the QPI signal due to the large spatially dependent modulation to the LDOS the ionization ring causes.

Figures 2(b) and (c) report the real space conductance maps taken on the same region as 2(a) acquired at bias voltage +0.70 V (tip current $I_t$ = 700 pA, AC excitation voltage $V_{osc}$ = 10 mV) and +0.90 V ($I_t$ = 850 pA, $V_{osc}$ = 10 mV) respectively. They reveal standing wave patterns in the electronic local density of states, which appears to be spatially modulated on a scale of a few nm, and the energy dependence of the QPI pattern is clearly visible, with the interference wavelength decreasing with increasing bias.

Figure 2(d) shows schematically how the Fermi wavevector **k** and QPI wavevector **q** in the (001) plane vary with energy $E$. The cone represents the dispersing band $E(\mathbf{k})$. For Na$_3$Bi we expect both conduction and valence bands to be nearly isotropic and circular, centred on the Γ point (k = 0) when projected on the (001) plane.[1] A constant-energy cut yields a circle of energy-dependent radius **k**. Intravalley scattering processes caused by disorder scatter the electrons from one point to another on a single constant-energy circle through a scattering wavevector **q**. This results in $|\mathbf{q}| \leq 2k$, with a cusp-like maximum at $|\mathbf{q}| = 2k$ due to the enhanced joint density of states (JDOS).

The 2D Fourier transforms of the conductance maps shown in Figures 2(e)-(f) convert the observed spatial oscillations to reciprocal space and reveal a ring of constant radius $2k$, agreeing with our assumption of an isotropic and circular dispersion $E(\mathbf{k})$ in the (001) plane. Figure 2(g) plots the radial average of the intensity of the 2D Fourier transforms from conductance maps taken at 0.45 V, 0.55 V, 0.65 V, 0.75 V and 0.85 V. It is clear that the dominant wavevector of the observed QPI $|\mathbf{q}| = 2k$ (the radius of the ring), visible as an increase in the radial intensity signal and marked by pink arrows as a guide for the eye, varies significantly as a function of the applied bias $V_b$. The feature around **q** = 0 shows no energy dependent variation, as it is the expected q = 0 peak corresponding to the local density of states that is broadened due to the finite resolution, and pointing out to the existence of a single band in this energy range above the Dirac node.

Figure 3 shows that similar quasiparticle interference features are also observed at negative bias. Figures 3(a)-(b) represent the d$I$/d$V$ maps acquired at $V_b$ = -0.20 V ($I_t$ = 250 pA, $V_{osc}$ = 10 mV) and -0.30 V ($I_t$ = 300pA, $V_{osc}$ = 10 mV) respectively, on the same area reported in Figure 2(a). Figures 3(c)-(d) are the corresponding 2D Fourier transforms of (a)-(b) respectively; Figure 3(g) is the radial average of 2D Fourier transforms from conductance maps taken at -0.20 V, -0.25 V, -0.30 V, -0.35 V, -0.40 V, -0.55 V and -0.65 V. The real space modulation of the local density of states is clearly visible at negative biases too, with the typical scattering wavelength becoming shorter with larger applied bias. The corresponding Fourier transforms show that the quasiparticle scattering on the (001) plane is to a good approximation isotropic also below the Dirac point, as in reciprocal space we observe ring-like structures compatible with multiple cone-like dispersions. For negative biases, indeed, we can identify more than one dispersing feature, as highlighted by the arrows of different colours included in Fig. 3(e) as a guide for the eye. Additional features are also observed around **q** = 0, with a clear energy dependent variation in the size of the feature with energy. This suggests scattering processes from multiple bands in the valence band as discussed below.

To understand this further and to compare with theory calculations we fit the peaks corresponding to $2k$ in Fig. 2(g) and Fig. 3(e) with Lorentzian peaks overlaid on a Log-cubic background (to account for the zero-q peak in QPI[19,22]). Exemplary peak fittings are shown in Figure 3(f) for the valence (top panel) and conduction bands (bottom panel). Additional raw spectra with the corresponding peak fitting and FFTs are reported in figure S1.[36]

Figure 4(a) shows the electron tunnel energy $E = eV_b$ versus $k = q/2$ for all the QPI features (markers). The QPI features indicate a single linearly dispersing electron-like band above the Fermi energy ($E > 0$) and three linearly dispersing hole-like bands below the Fermi energy. This agrees with the expected low-energy bands which have character $|P_{\frac{3}{2}}^-, \pm\frac{3}{2}>, |P_+^-, \pm\frac{1}{2}>, |P_{\frac{3}{2}}^+, \pm\frac{3}{2}>, |S_{\frac{1}{2}}^-, \pm\frac{1}{2}>$ at the $\Gamma$ point,[3] and with the relative intensity of conduction and valence band observed via STS [see Figure 1(b)]. We fit each dispersion to a linear relation $|E - E_0| = \hbar v_F k$ where $\hbar$ is Planck's constant, $v_F$ is the band velocity, and $E_0$ a fit parameter. The linear fittings are superimposed as solid lines on the experimental data. We obtain $v_F = 1.6 \times 10^6 \ ms^{-1}$ for the electron-like band above the Fermi energy (pink dots); while the three bands observed below the Fermi energy have band velocities of $v_F = 4.2 \times 10^5 \ ms^{-1}$ (green squares), $v_F = 6.7 \times 10^5 \ ms^{-1}$ (orange triangles) and $v_F = 1.6 \times 10^6 \ ms^{-1}$ (blue diamonds) respectively.

We compare our experimental results with first-principles theoretical calculations using the generalized gradient approximation (GGA); SCAN, an improved meta-GGA method; Heyd–Scuseria–Ernzerhof exchange-correlation functional (HSE06) and the GW approach. In Figure 4(a) we overlay the band structure obtained by GGA and GW for bulk Na$_3$Bi along the $\tilde{\Gamma}$–$\tilde{K}$ direction (in the plane containing the Dirac point) to the experimental data; in Figure 4(b) we show the electron band velocities extracted from calculations and experiment. Figure 4(c) shows the comparison of computational results obtained by using the GGA (solid blue lines) and GW (dashed red lines) methods over the entire Brillouin zone. Full results obtained via GGA, meta-GGA, HSE06 and GW methods are reported in Figure S2.[36] DFT calculations qualitatively agree with the experimental findings, showing a single conical electron-like band above the Fermi energy, and three nested hole-like conical bands of increasing steepness (progressively higher velocities) for holes below the Fermi energy.

We observe a striking dependence of the band velocities on the method of approximation used in the *ab initio* calculation. For example, as shown in Fig. 4(b), the velocity of the electron-like band is underestimated compared to experiment by 69% in GGA, 65% in SCAN, 56% in HSE06, but only 18% in GW. The velocities follow a hierarchy which appears to correspond to the level of approximation used in accounting for interactions. The SCAN meta-GGA functional satisfies all known constraints and thus can be extremely accurate at a semi local XC level, while the HSE06 method has a more sophisticated and typically more accurate treatment of the XC potential, suggesting that it plays an important role in determining the overall band structure of the system. The GW approximation, which has a more accurate description of the XC potentials compared to DFT, shows even better agreement between theory and experiment for the conduction band; it predicts a Fermi velocity for holes $v_F =$

$1.3x10^6\ ms^{-1}$, close to the experimentally determined $v_F = 1.6x10^6\ ms^{-1}$. The disagreements between experiment and theory are largest for both the electron-like band and the deepest of the three bands below the Dirac cone. We attribute this to the different orbital character of these bands. In line with recent reports for few-layer Na₃Bi[10] and as evidenced in Figure S3 and S4,[36] in the [-2; 2] eV range around the Γ point, the first two bands below the Dirac node exhibit a strong Bi-*6p* character (indicated in blue in Fig. S3), while the electron-like band and the deepest of the three valence bands have mixed contribution from Na-*3s* and Bi-*s/p_z* orbitals (red in Fig. S3). We expect interaction effects to play a stronger role in determining the dispersion of this band, hence the larger disagreement with the experimental trend.

Despite the better agreement obtained with the GW method, we observe some significant discrepancies between the model and experiment: as seen in Figure 4(a), while preserving the asymmetry between the velocity of electrons and holes, GW only predicts two (nested) valence bands in the experimentally measured window, with the onset of the next valence band being about 1 eV below the Fermi level. However, the Fermi velocity we obtain experimentally for the topmost valence band ($v_F = 4.2x10^5\ ms^{-1}$) is consistent with previous ARPES reports on the (001) surface Na₃Bi bulk crystals and thick films,[1,13,15] which corroborates our QPI findings. We do not currently understand the origin of the discrepancy. One possibility is that this band originates from Na vacancies: it has been demonstrated[1] that with the increase of the surface potential, generated by the presence of Na vacancies creating an electric field on the surface, the surface state of Na₃Bi was gradually pushed up and away from the bulk states. This resulted in a cone-like dispersion along the external envelope of the bulk valence band. We also do not fully understand the origin of the strong interaction corrections to the band structure of Na₃Bi. Naively we would anticipate that interaction effect in Na₃Bi would be small, as the dimensionless interaction strength is α ~ 0.1 in this material;[5] further work is needed.

The conduction band of Na₃Bi has not been probed in ARPES which typically can only probe filled states below the Fermi energy. However, we note that there is some indirect evidence for a significantly higher electron band velocity, which supports our findings. A three-fold enhancement of the Fermi velocity of electrons compared to holes was reported in [14] from STS of Landau levels, however the measured electron velocity reported ($v_F = 7.9\ x10^5\ ms^{-1}$) is considerably lower than our findings, and was extrapolated from measurements in a much lower energy regime (which may not be above Lifshitz point). A high value for the Fermi velocity of the electrons ($v_F = 1x10^6\ ms^{-1}$) was observed in ARPES on ultrathin films of Na₃Bi,[11] where, as in our work, the experimental value was much higher than theoretically predicted.

CONCLUSIONS

Here we present low-temperature scanning tunnelling microscopy and spectroscopy measurements of (60 x 60) nm² large, atomically smooth areas of Na₃Bi(001) thin films, grown on Si(111) via MBE. We perform quasiparticle interference spectroscopy, and observe (concentrical) ring-like structures in the scattering wavevector well above (below) the Dirac nodes, whose diameter scales linearly with the applied voltage, which we use to map the energy dependent Fermi wavevector of the conduction and valence bands. The extracted Fermi velocities for the lowest energy hole band $v_F = 4.2 \, x10^5 \, ms^{-1}$ is in good agreement with previous observations, and the electron band velocity $v_F = 1.6 \, x10^6 \, ms^{-1}$ is qualitatively consistent with observations of a higher velocity for electrons compared to holes, and also may explain the very high mobility in highly n-doped Na₃Bi. We compare these experimental findings to the theoretical band structure calculated both DFT and GW method: the agreement between experiment and theory significantly improves with more refined approaches to exchange and interaction potential in DFT, while GW calculations give an electron velocity that is within 18% of the experimental value, suggesting exchange-correlation effects to be the cause of the observed discrepancy for this system.

Acknowledgments

M.T.E. was supported by ARC DECRA fellowship DE160101157. I. D. B, M.T.E., J.L.C., and M.S.F. were supported by ARC CE170100039. W.W. and S.A.Y. were supported by Singapore MOE AcRF Tier 2 (MOE2019-T2-1-001). The computational work was performed on resources of the National Supercomputing Centre, Singapore and the Texas Advanced Computing Center.

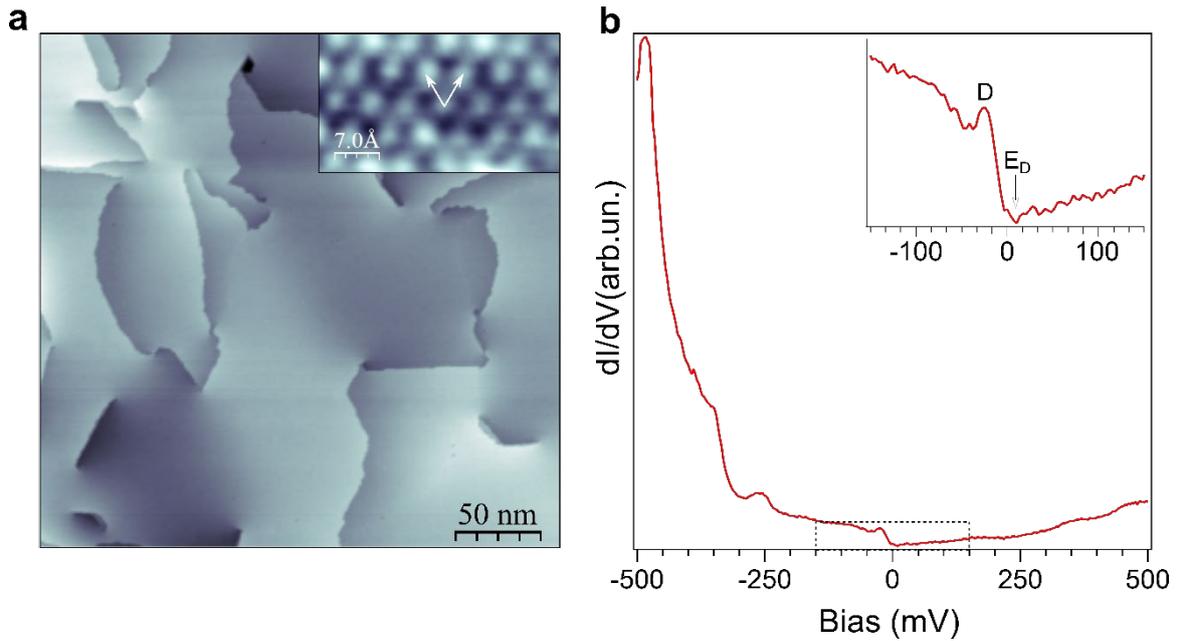

**Figure 1**. Topographic and spectroscopic characterization of Na$_3$Bi thin film. (a) 300 nm x 300 nm topographic STM image (bias voltage $V$ = -3 V and tunnel current $I$ = 50 pA) showing the overall topography of the sample; inset: atomically resolved image taken on a separate 20 nm Na$_3$Bi film, where the white arrows mark the 5.45Å lattice constant. (b) Point scanning tunnelling spectrum dI/dV vs bias on 20 nm Na$_3$Bi film. (Bias modulation rms amplitude is 5 mV). The bias voltage position of minimum d$I$/d$V$, visible in the inset, is labelled E$_D$ and identified as the Dirac point energy.

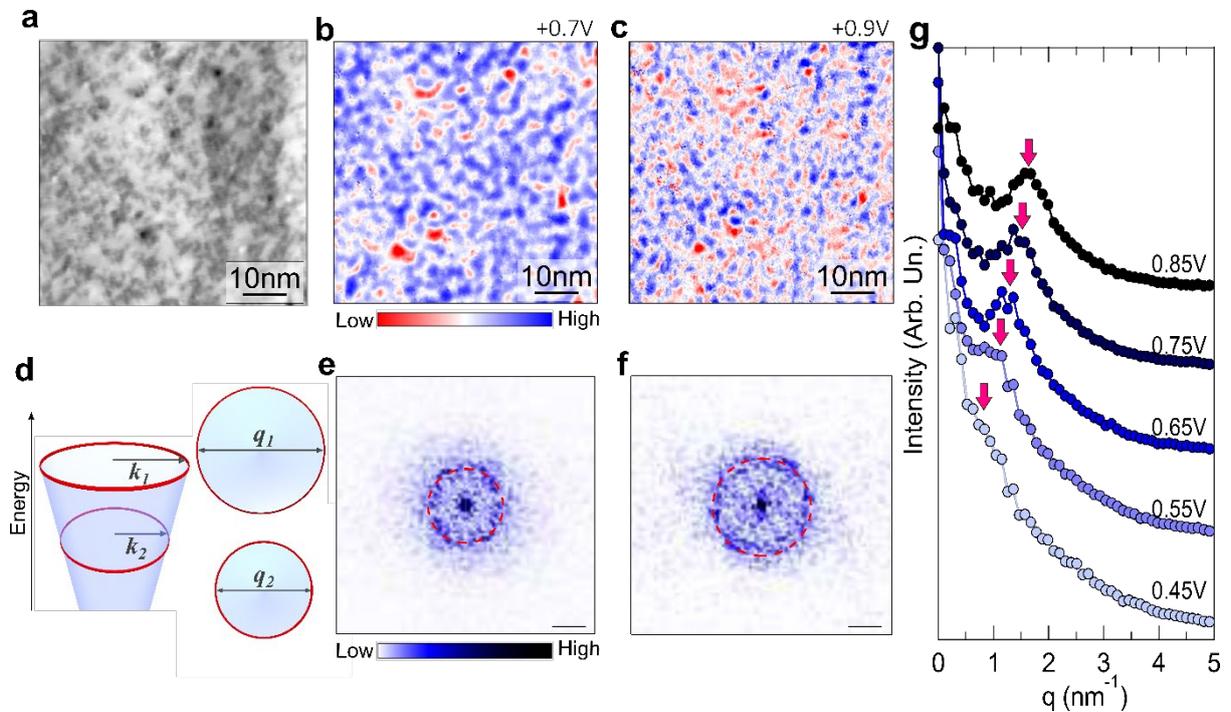

**Figure 2**. Topography and quasiparticle interference mapping of the conduction band of a Na$_3$Bi thin film. (a) 60nm x 60 nm topographic STM image (bias voltage $V$ = 0.375 V and tunnel current $I$ = 600 pA) showing an atomically flat region of 20 nm Na$_3$Bi on Si(111). (b)-(c) d$I$/d$V$ maps of the same area

as shown in Fig. 1(a) obtained at $V$= 0.7 and 0.9V, and $I$ = 700, 850 pA respectively. (d) 3D Schematic of the binding energy vs. $\mathbf{k_x}$ - $\mathbf{k_y}$ on the (001) plane of Na$_3$Bi, with the red circle representing a constant-energy contour for states around the Γ point, and the intravalley wavevector scattering process denoted by $\mathbf{q}$. (e)-(f) Two-dimensional Fourier transform (2D FT) of (b)-(c) respectively where the scale bar represents 2.1 nm$^{-1}$. (g) Radial averaged intensity profiles of the 2D FT acquired at $V$= 0.45 V, 0.55 V, 0.65 V, 0.75 V, 0.85 V (2D FT not shown) plotted as a function of wavevector $q$. Curves are offset for clarity. Pink arrows denote the position of the main dispersing feature as a guide for the eye.

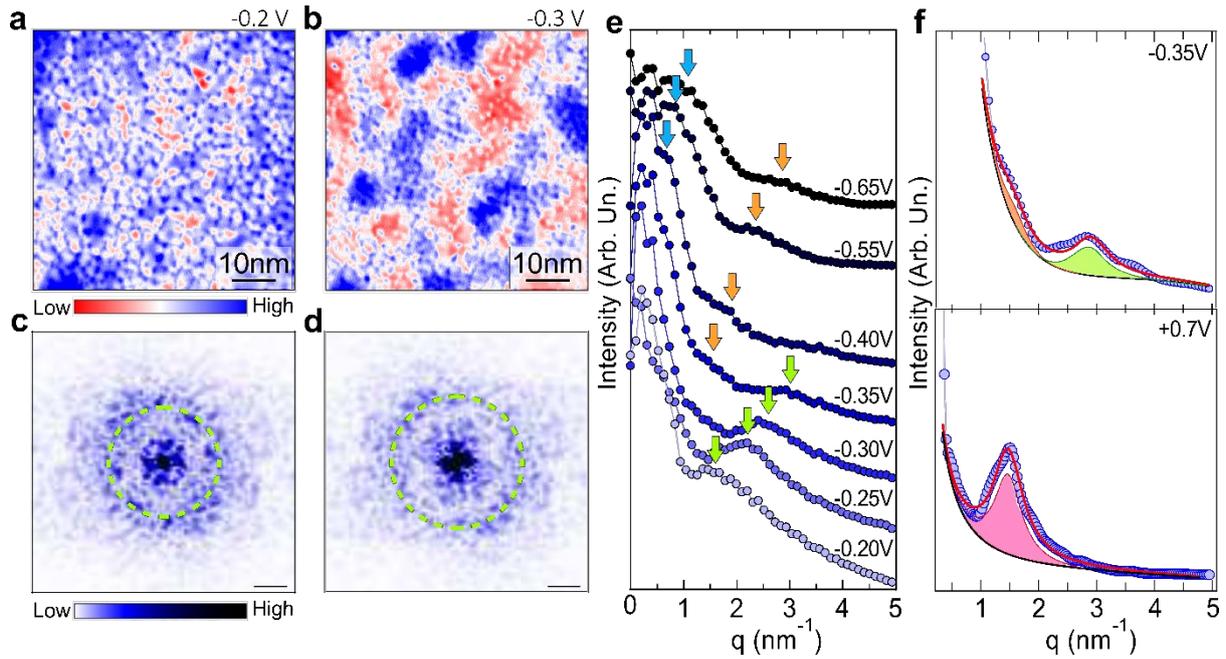

**Figure 3**. Quasiparticle interference mapping of the valence bands of a Na$_3$Bi thin film. (a)-(b) d$I$/d$V$ maps of the same area as shown in Fig. 1(a) obtained at $V$ = -0.25 V and -0.30 V respectively; $I$ = 450 pA in both cases. (c)-(d) Two-dimensional Fourier transform (2D FT) of (a)-(b) respectively where the scale bar represents 2.1 nm$^{-1}$. (e) Radial averaged intensity profiles of the 2D FT shown in (d)-(f) (along with -0.25 V, -0.35 V, -0.4 V, -0.55V and -0.65 V, 2D FT not shown) plotted as a function of wavevector $q$. Curves are offset for clarity, vertical arrows follow the dispersion of the main features as a guide for the eye. (f) Exemplary peak fitting for the radial averaged intensity profile of the 2D FT for the mappings collected at $V$ = -0.35 V (top panel) and $V$ = 0.7 V (bottom panel).

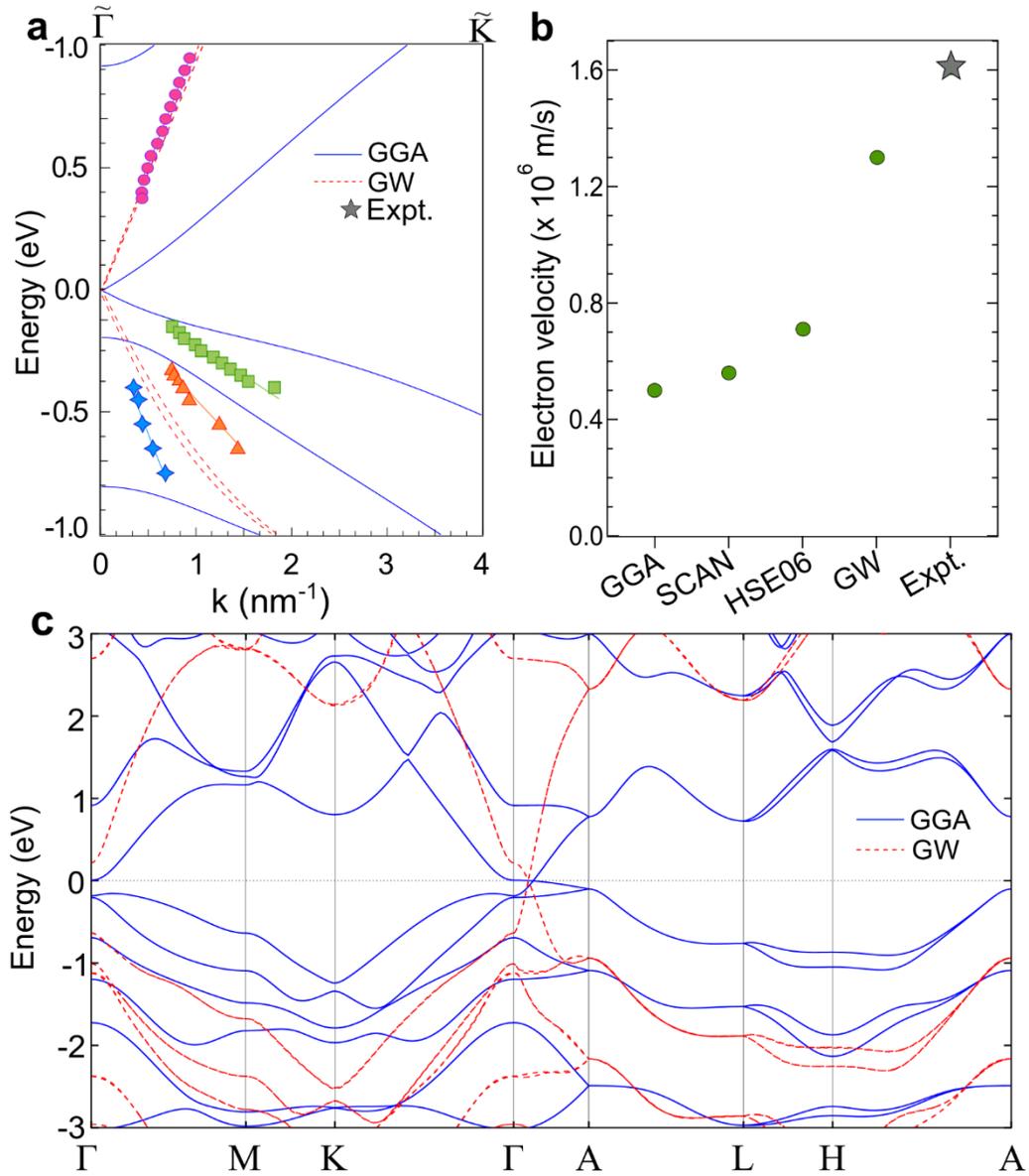

**Figure 4**. Comparison between experimental results and calculation for the band structure of Na$_3$Bi. (a) Calculated band structure for Na$_3$Bi along the $\tilde{\Gamma} - \tilde{K}$ direction in the plane containing the Dirac point. Markers are the measured data from QPI, linear fittings to evaluate the Fermi velocities are superimposed on the data. (b) Electron velocity evaluated with DFT and GW methods (green dots) and experiment (grey star). In each case the velocity is obtained from a linear fit to $E(k)$ over the range 450 meV $\leq E \leq$ 950 meV. (c) Comparison of the calculated band structures for Na$_3$Bi with GGA (solid blue lines) and GW (red dashed lines) approaches along the main symmetry directions.

# Supplemental Material: Band Dispersions in Topological Dirac Semimetal Na$_3$Bi from Quasi-particle Interference

I. Di Bernardo,[1,2] J. Collins,[1,2] W. Wu,[3] Ju Zhou,[4] Shengyuan A. Yang,[3] Ju Sheng,[4] M. T. Edmonds,[1,2,5]* M. S. Fuhrer[1,2,5]*

[1] Australian Research Council Centre of Excellence in Future Low-Energy Electronics Technologies, Monash University, 3800 Clayton, Victoria, Australia

[2] School of Physics and Astronomy, Monash University, 3800 Clayton, Victoria, Australia

[3] Research Laboratory for Quantum Materials, Singapore University of Technology and Design, Singapore

[4] School of Physical Science and Technology, Soochow University, Suzhou, 215006, China

[5] Monash Centre for Atomically Thin Materials, Monash University, 3800 Clayton, Victoria, Australia

* Corresponding authors: mark.edmonds@monash.edu and michael.fuhrer@monash.edu

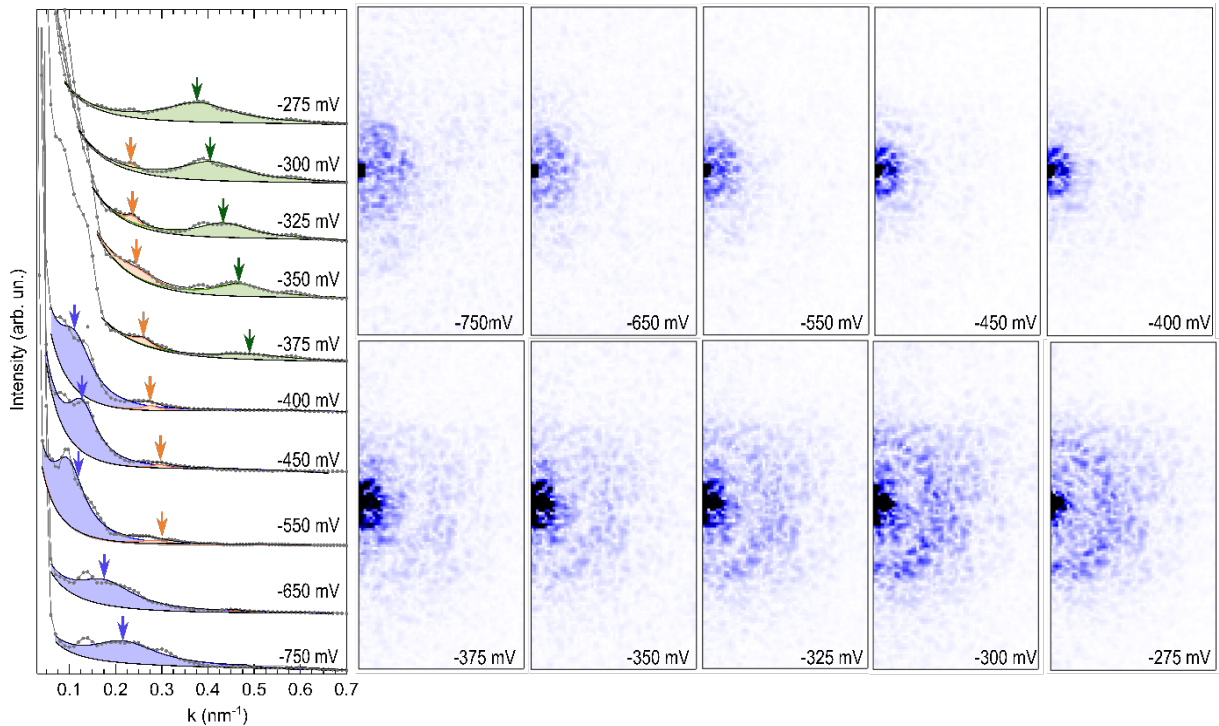

**Figure S1**: Additional raw data and peak fitting for the radially averaged QPI data at negative voltages, along with the corresponding FFTs.

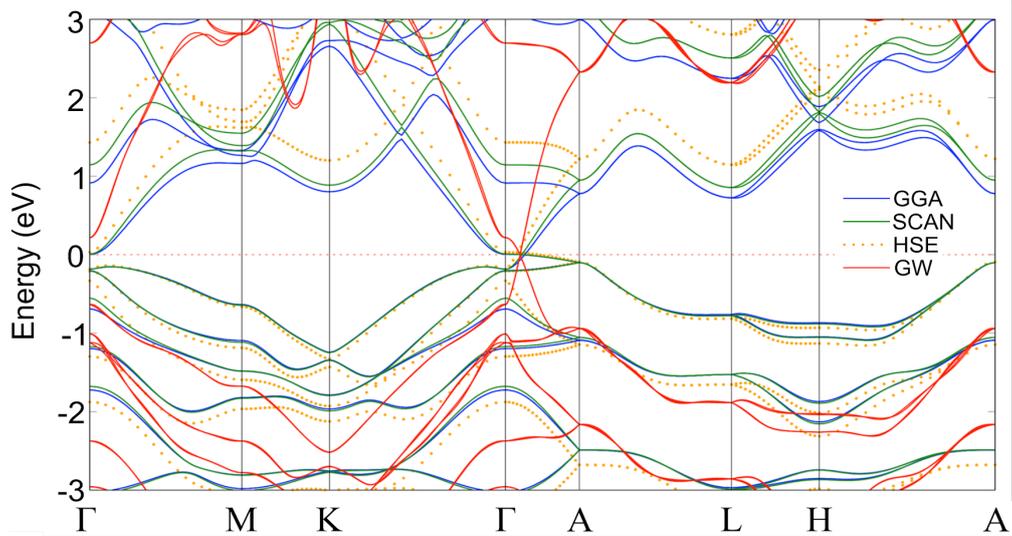

**Figure S2**: Band structure calculations with GGA, meta-GGA, HSE06 and GW approaches.

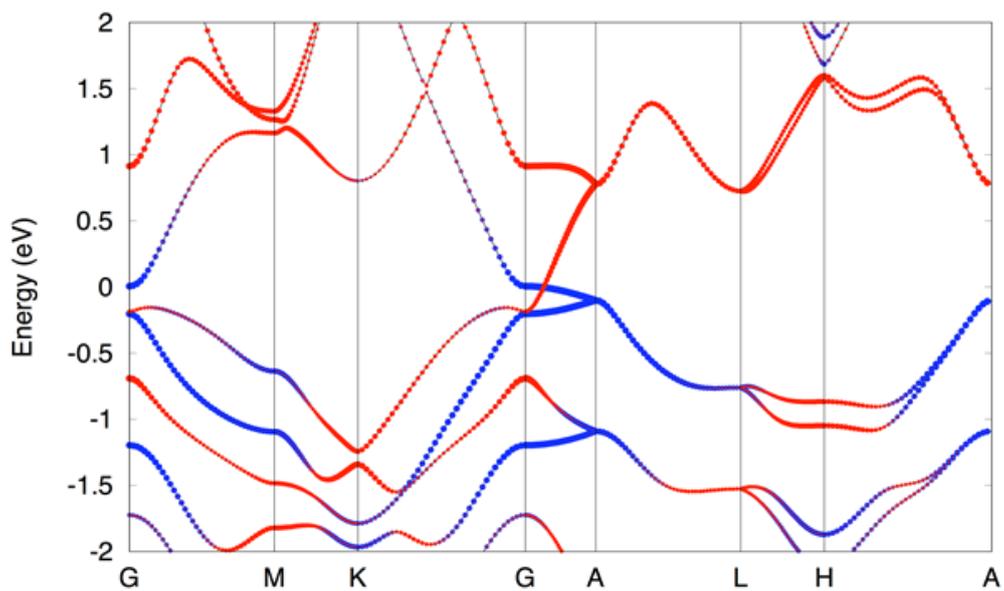

**Figure S3**: Orbital contributions to the bandstructure of $Na_3Bi$, calculated via GGA method. Contributions from the $s/p_z$ orbitals of Na and Bi atoms, and the $p_x/p_y$ orbitals of Bi atoms are denoted by red and blue dots, respectively.

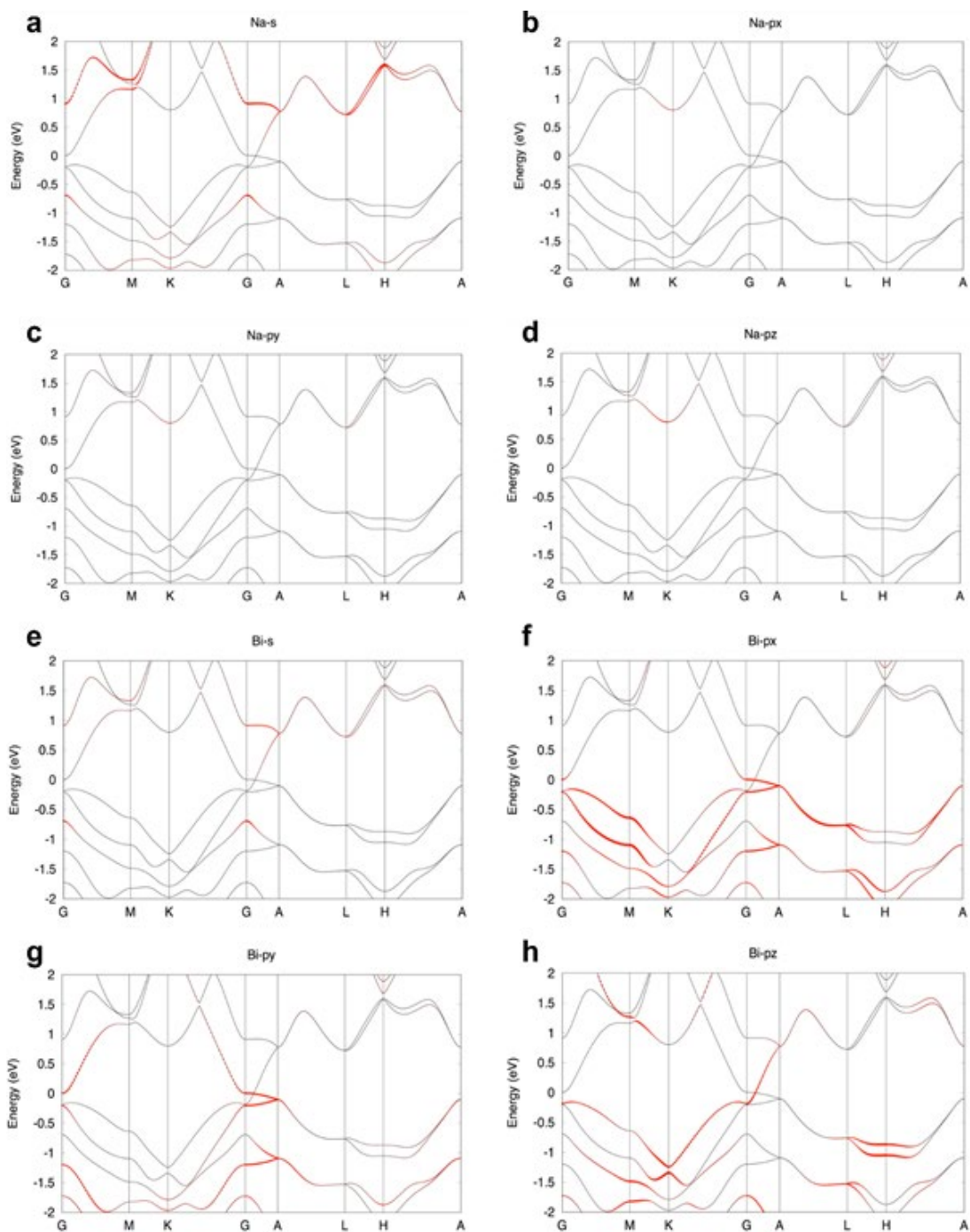

**Figure S4**: Contributions of the single atomic orbitals to the bandstructure of $Na_3Bi$, denoted by red dots. Panels (a)-(d) report the data for Na atoms, panels (e)-(h) the data for Bi atoms.